Paths to Influence – How Coordinated Influence Operations Affect the Prominence of Ideas

Darren Linvill and Patrick Warren (Clemson University)

**I. Intro**

Malicious, inauthentic digital content has, in recent years, become an issue of increased concern. The online ecosystem is used by foreign provocateurs to influence elections (Linvill & Warren, 2020), by autocrats to control domestic discourse (Martin et al., 2019), and by fraudsters who are making off with more of their victims' money with each passing year (Lyngaas & Rabinowitz, 2023). The ways in which bad actors work to engage with their audiences, however, can too often be narrowly conceived. Troll farms are not all run the same way, and their tactics and strategies are shaped in important ways by their goals and by their capabilities.

Even if we take what is perhaps the most famous social media coordinated influence operation (CIO) to date, the Russian Internet Research Agency's (IRA) efforts to influence the 2016 U.S. election, we can see multiple approaches. While the IRA did employ accounts which purported to be genuine Americans of various political beliefs, what may now be considered the prototypical troll accounts, they conducted more specialized activity as well. Among these were at least 55 newsfeed accounts identified by Linvill and Warren (2020). These accounts each purported to be news aggregators, most from a specific U.S. city. They appeared professional with well designed profiles giving the appearance of credibility. These "Newsfeeds" almost exclusively posted real news they gathered from genuine local sources. Their efforts were of course not altruistic in nature, however. Ehrett et al. (2022) found that these accounts were attempting to serve an agenda-setting function. They shared stories which portrayed the world as a dangerous place, especially for vulnerable minority groups, at rates far greater than did the outlets from which they pulled their stories. The stories were real, but their prominence was not, and in this way they fanned the flames of disillusionment and discontent prior to the election.

But online fraud and disinformation is not always farms of trolls diligently pretending to be your neighbor or local news outlet. The work of the IRA in 2016 and after can be juxtaposed to ongoing efforts affiliated with the People's Republic of China (PRC). In October, 2019, Daryl Morey, then general manager of the National Basketball Association's Houston Rockets, tweeted an image saying "Fight for Freedom. Stand with Hong Kong." The Wall Street Journal (Cohen et al., 2019) reported that in the week following Morey's tweet, he was flooded with replies from thousands of PRC affiliated troll accounts. These accounts attacked Morey personally, but also pushed narratives about Chinese sovereignty and American hypocrisy. In the wake of these attacks, Morey apologized for his tweet.

In these two cases, the goals of Russian and Chinese affiliated accounts were opposite. The IRA worked to highlight narratives they preferred. China worked to suppress narratives they did not.

These examples, pulled from genuine Russian and Chinese influence operations, illustrate the breadth of tactics bad actors may employ through social media to influence people for nefarious ends. Tactics vary depending on the goals and constraints of the actor. The objective of this paper is to build a model capturing how inauthentic influence operations are conducted and why they target specific conversations using the tactics they do. First, however, we will illustrate the set of tactics operations have to choose from through specific, real-world examples.

## II. Many Paths to Influence

Social media influence operations may have a variety of specific goals. They can, for instance, attempt to espouse a particular narrative or world view, engage in reputational management, motivate group action, persuade users of a particular argument, or sow chaos and division. To simplify the list of potential goals, however, we will start with the generalization that social media influence operations work to affect the prominence of ideas. There are, however, myriad tactics one might employ to engage in such influence, and the ideal choice of tactic depends on a variety of factors. As we discussed in the introduction, depending on both context and goals, two social media influence operations may appear staggeringly different from one another and yet each be utilizing the best tactics available given the actors' constraints.

Influence-operation tactics all fall into one of four relatively broad categories dependent on two simple factors. Actors choose which category of tactics they employ, but constraints may limit reasonable options and push an operation into employing a particular tactic set. The first factor is the actor's goal; is an operation working to either promote or demote a focal idea relative to the organic prominence it would have absent the work of the operation. An actor may desire an idea be more salient (promote) or less salient (demote) in the minds of the public. The second factor addresses the mechanism by which the goal is targeted. What ideas are best exploited in order to influence the salience of the focal idea will vary but there are two basic choices, one can employ a direct mechanism and exploit the focal idea itself or one can employ an indirect mechanism and exploit one or more ideas that are not the focal idea. Depending on the goal (promote or demote) and the mechanism (direct or indirect), any operation may be placed in at least one of four blocks in a matrix (see Figure 1). We will explain each of these tactical categories below, by example.

Figure 1. Influence Operation Tactics Matrix

| | Direct Mechanism | Indirect Mechanism |
|---|---|---|
| Goal: Promote Focal Idea | Promote Focal Idea by Strengthening Focal Idea | Promote Focal Idea by Weakening Alternative Idea(s) |
| Goal: Demote Focal Idea | Demote Focal Idea by Weakening Focal Idea | Demote Focal Idea by Strengthening Alternative Ideas(s) |

**Example Tactic #1: Directly Promote Focal Idea**

The IRA Troll farm first rose to prominence following a 2015 New York Times report about their activity (Chen, 2015). They became infamous, however, due to their attempts to influence the 2016 U.S. Presidential election. But before they were active in U.S. politics, they operated in Russian—targeting a domestic audience and Russian-speakers in neighboring Ukraine. Though the organizational structure of this persistent operator is murky and appears to have changed over time, the IRA or a successor organization has continued to operate up to, at least, the summer of 2023 and may continue today.

Throughout its operation the IRA has used myriad strategies and methods of influence. But the method they are best known for is the creation and cultivation of deep and complex identities, both individual and organizational, with the goal of promoting particular ideas. In the context of the infamous campaign targeting the U.S. in 2016, these identities fell into one of several specific types (Linvill & Warren, 2020), including the newsfeed accounts previously discussed but also accounts that purported to be individuals with identities and messaging focused on either specifically right leaning or left leaning ideologies. These included numerous prominent accounts with tens and even hundreds of thousands of followers. Some of these persona were deep, with accounts on multiple platforms and quotes attributed to them across mainstream media (Xia et al., 2019).

But despite a degree of bespoke heterogeneity, accounts shared several common characteristics. IRA persona engaged in messaging consistent with extreme, partisan versions of these ideological groups and attacked moderate perspectives. The IRA operation attacked centrist and institutionalist world views from both the left and the right simultaneously. They questioned democratic processes, scientific consensus, and America's place on the global stage. The IRA worked to promote an extreme view of the world, one which they evidently hoped would spread disillusionment and discontent among those that engaged with the messaging and accrue indirect benefit to the Kremlin.

The divisive ideas IRA trolls worked to promote are clear in the content they shared (Linvill & Warren, 2020). One of the most used hashtags from their right leaning Twitter persona included #IslamKills and they targeted right leaning users with messages such as "#ThanksObama We're FINALLY evicting Obama. Now Donald Trump will bring back jobs for the lazy ass Obamacare recipients." Similarly, a top hashtag from left leaning IRA Twitter persona was #PoliceBrutality and they targeted left leaning users with messages such as "NO LIVES MATTER TO HILLARY CLINTON. ONLY VOTES MATTER TO HILLARY CLINTON." These ideas were already circulating throughout the social media ecosystem; it was a tactical choice for the IRA to raise the prominence of these ideas through direct promotion. There were no constraints posed by the ideas themselves beyond the ability of Russian operators to convincingly engage with American users, and these difficulties were overcome by a well resourced operation.

**Example Tactic #2: Indirectly Promote Focal Idea**

In March, 2022, Russia's full scale invasion of Ukraine was faltering and so was their related propaganda war. Footage of Russian losses (e.g., Sabbagh, 2022) was damaging perceptions of Russian military superiority and calling into question Russian leadership.The Russian disinformation and propaganda machine needed some positive messaging to bolster the homefront and ensure support for the Kremlin. For Russia, however, there was very little good news coming out of the conflict in those early weeks. An expected easy victory had not transpired, Kyiv had not fallen, and losses were mounting. Something had to be done.

Without good news to promote the war and Russian dominance, the Kremlin turned instead to attacking information about Ukrainian successes. This mostly involved undermining Western media and Ukrainian claims of battlefield success. One form these efforts took was in posts from inauthentic social media accounts believed to be affiliated with the Russian IRA (Kao & Silverman, 2022a). These accounts, engaging in Russian, shared cartoons and other satirical messages lampooning Western narratives about the course of the invasion. They also shared stories that were clearly fake. One video shared by these accounts claimed to show a German broadcast of a journalist standing in front of what was purported to be dozens of civilian body bags in Kyiv. In the middle of the broadcast one of the "bodies" sits up. The troll accounts claimed the video proved the West was spreading fake news about casualties when, in fact, the video was from a protest in Austria against global climate change and the body bags contained live protestors.

Perhaps yet more malicious, these and other Russian affiliated accounts also shared fake fact-checks of Ukrainian claims (Kao & Silverman, 2022b). These posts shared juxtaposed images. One image would show an "original", easily found from an internet search. These included images of Russian vehicles or Ukrainian cities. The other would show a Ukrainian "fake," an altered version of the first image to show the Russian vehicle burning or the Ukrainian city being shelled (See Figure 2). These fact checks would suggest that Ukraine was responsible for the altered images and that the Ukrainians were spreading disinformation. Analysis of image meta-data, however, showed these fake fact-checks were created at the same moment by the same person as the alleged "Ukrainian fake," which never previously existed and did not spread independently.

Russian attempts to undermine Western and Ukrainian narratives about the war employed a common disinformation tactic–the Firehose of Falsehood– using vast quantities of false or ambiguous claims to undermine users' beliefs in any objective truth (Paul & Matthews, 2016).

Russian influence operations may not have been able to convince their audience that the war was going well, but it was perhaps sufficient to level the propaganda playing field and persuade their audience to believe nothing. By diminishing other narratives they raised at least the relative prominence of their target ideas.

Figure 2. Example screenshot of Russian fake fact check video (left), with English translation (right)

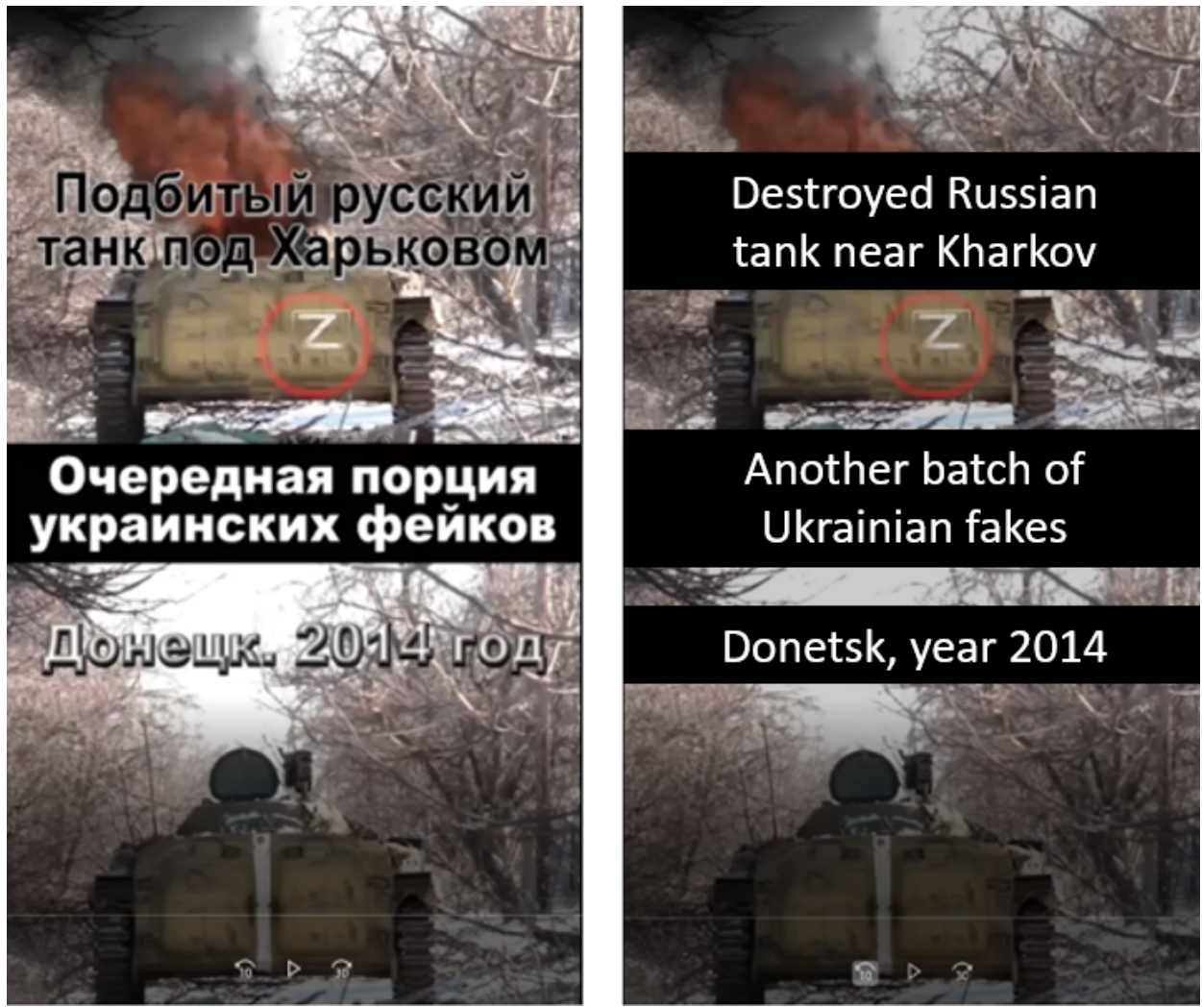

**Example Tactic #3: Directly Demote Focal Idea**

China hosted the Beijing Winter Olympics in February 2022. As with any host nation, China used the games as a promotional opportunity. Some human rights activists, however, saw the games as a moment to raise awareness of Chinese atrocities, especially those committed against China's Uyghur Muslim minority.

Human rights organizations have accused China of crimes against humanity targeting the Uyghur, a Muslim minority from north-western China's Xinjiang region. Evidence exists of crimes including forced sterilization, forced labor, and forced detention of Uyghur Muslims (Maizland, 2022). Prior to the 2022 Beijing Olympics, activists began using #GenocideGames to organize conversations on social media. The hashtag was intended to directly link conversations about the Olympics with conversations about atrocities committed against the Uyghur.

A report from the Wall Street Journal found PRC affiliated Twitter accounts flooded the platform with tens of thousands of posts using #GenocideGames prior to the Beijing games (Wells & Lin, 2022). Counterintuitively, it was claimed Chinese affiliated accounts were using the hashtag in an attempt to weaken the usefulness of the hashtag and silence their critics. By using the hashtag on messages that were entirely unrelated to discussions of Uyghur atrocities, China made it incrementally less likely for authentic users wanting to engage in those conversations to find what they were looking for. In using clearly inauthentic accounts, China may have also been working to influence Twitter's algorithm into suppressing these same conversations.

China faced a variety of constraints in deciding how to address human rights related attacks while keeping the focus on a successful Olympic Games. It was difficult to engage directly with accusations regarding the Uyghur, there was ample evidence of China's abuses and engaging with the conversations directly could serve to simply spotlight them. Without the option to change hearts and minds, flooding the hashtag and diluting the discourse was a logical tactic.

**Example Tactic #4: Indirectly Demote Focal Idea**

In September, 2021, Chinese virologist Li-Meng Yan published a report alleging the Covid-19 virus was artificially created in a Chinese laboratory (Timberg, 2021). Though the report was discredited at the time by prominent researchers, this narrative remained a clear threat to China. It was well established that Covid-19 originated in Wuhan, China, but prior to Li-Meng Yan's report, little evidence had been presented which pointed blame at the Chinese government for its creation.

Impacting conversations about Covid-19, however, would have been extremely difficult. In September 2021 social media was awash with users discussing the virus, the pandemic remained the biggest news item of the year.[1] Attempts to discredit Li-meng Yan's story about Covid-19's origins would need to focus on discrediting her and not on influencing wider conversations about the pandemic.

A report from the Australian Strategic Policy Institute (ASPI) found tens of thousands of tweets posted between April and June, 2021, which they linked to a Chinese state operation (Zhang, 2021). The messages attacked the credibility of Yan, as well as the Chinese dissident billionaire Guo Wengui and the Republican strategist Steve Bannon, both of whom worked to promote

---

[1] There is some evidence that China attempted to create a narrative that Covid-19 originated from frozen Maine lobster which were shipped to Wuhan, but this attempt was not significant or successful (Solon et al., 2021).

Yan's work. Importantly, all of the tweets identified in the report used the hashtags #StopAsianHate along with #LiMengYan. The posts employed the same memes and messaging which suggested Yan's report blaming China for the virus was anti-Asian discrimination.

Conversation online about racism targeting Asians were particularly prominent in this period of the pandemic and employing #StopAsianHate was an attempt to co-opt these discussions and link them to Yan. The ASPI report found content from this campaign (see Figure 3) across thousands of accounts and multiple online platforms. Co-opting existing narratives to discredit Yan and her report as a form of racism was a reasonable tactic given the constraints any operations would have faced trying to steer the much larger discourse which existed around the origins of Covid-19.

Figure 3. Memes shared by Chinese disinformation campaign attacking Li-Meng Yan.

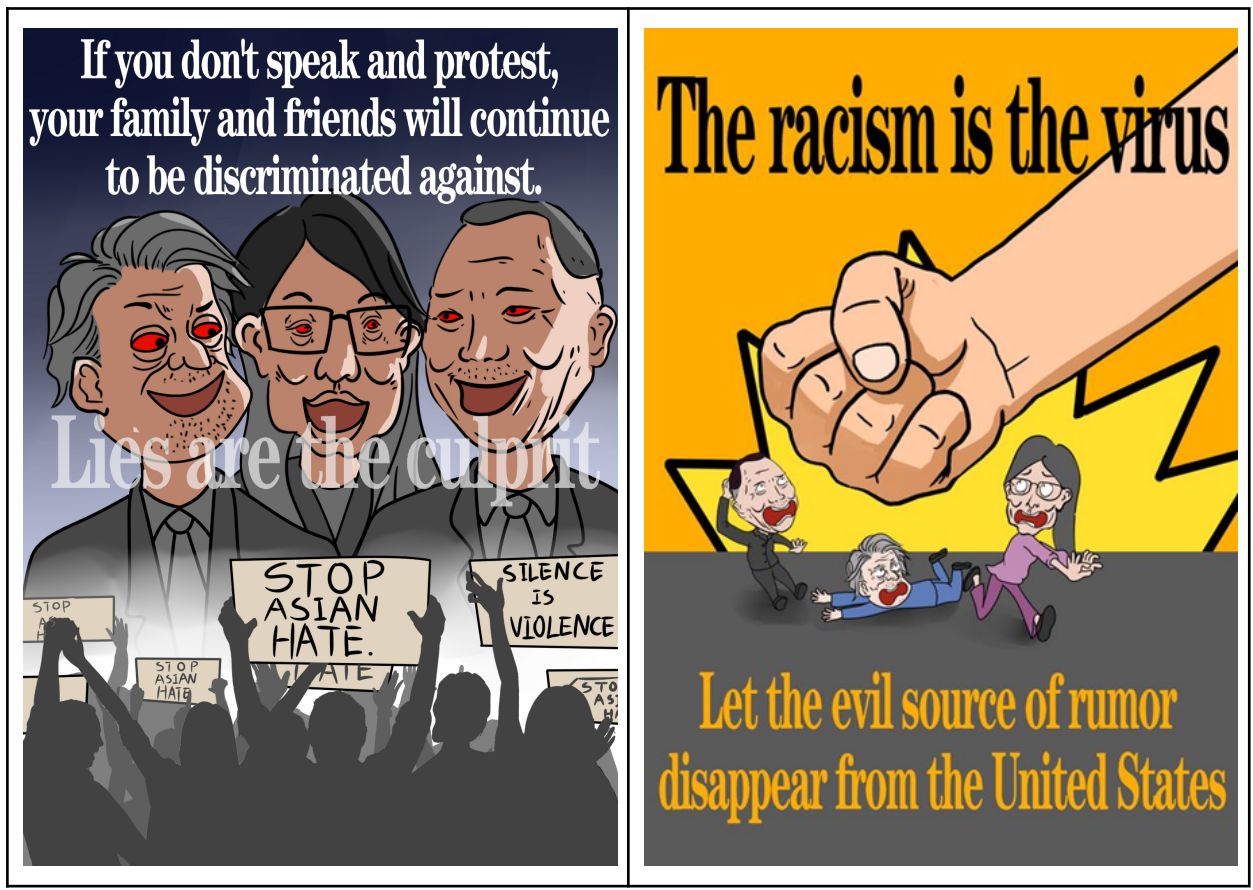

### III. A (mostly) economic model of CIO behavior

We present, in this section, an economic model of the productive decisions of a Coordinated Influence Operation (CIO). In it, the CIO decides what ideas to target for strengthening or weakening, and how much of what inputs to use to exert that influence. This model borrows much from the standard economic model of profit-maximization by a multiple-product monopolist who produces a variety of related products (Forbes, 1988).

The input side is straightforward, where for whatever output profile the producer chooses he will choose the input mix that minimizes his costs of producing that output. Those inputs could come in a variety of forms: paid staff, computing power, AI-generated posts, bots for hire, advertisements, media, and a huge variety of other sources of influence. Holding output fixed, the mix of inputs will depend on their relative prices and productivities. Differences in circumstance that shift the level or mix of outputs will only affect input mixes to the extent that they might affect those relative productivities. For example, if one of the inputs (what you might call the high-quality input) is relatively advantageous when producing high levels of influence, we would predict to see that input used more intensely when the producer has particularly strong incentives to influence a narrative. Thus, we might expect mostly low-quality inputs (cheap and simple bots, substantial reuse of content) when the desired level of influence is low but substantial use of high-quality inputs (human-curated content, cyber-enabled influence, original media) when the desired level is high.

Optimal influence output choices for each idea will depend on these costs of production, but also on the value to the producer of a marginal shift in the prominence of that idea and the patterns of substitution amongst alternative ideas. All else equal, we'd expect more influence to be exerted when the prominence of an idea is more important, when influence efforts have a bigger impact on prominence, and when influence is inexpensive to produce.

This model provides some clear predictions for when we should expect CIOs to appear in each sector of our 2x2 schema of promotion/demotion and strengthening/weakening, mostly depending on their goals and the relative impact of influence. At one extreme, when the CIO has one or a handful of ideas in the topic that he is particularly interested in making prominent, we should typically see intense strengthening of those ideas, using high-quality inputs. The exception occurs when that idea is particularly difficult to successfully influence, relative to alternatives. In that case, the CIO might adopt a strategy of weakening all the alternatives, using low-quality inputs. At the opposite extreme, when the CIO has one or handful of narratives in the topic that he is particularly interested in making less prominent, the optimal strategy is more typically a mix of weakening the focal idea and strengthening alternatives, depending on the effectiveness of strengthening versus weakening.

### III. A. Formalizing the Model.

Consider a CIO who wants to influence the beliefs and/or behavior of some collection of consumers/citizens. Suppose, for simplicity, that there is a set of N ideas, indexed by i, that are competing for attention and support on some focal topic in an information space. The CIO can take some set of costly actions to influence the prominence of each of these narratives, but the impact of these actions depend on some characteristics of the ideas and of the CIO.

Formally, the operator can choose a 3-dimensional vector of investments $(x_i, q_i, d_i)$, where $x_i$ represents the quantity investment in promoting idea i, and $q_i$ represents the quality investment in promoting idea i, and $d_i \in \{1, -1\}$ represents the direction of influence, where the difference between quality and quantity is how their relative efficiencies in the production of influence change with the scale of influence. These investments come with constant marginal costs $p_x > 0$ and $p_q > 0$, which are independent of idea but might vary across topics or influence actors.

The impact of these investments on the strength of an idea is given by $a_i d(1 + \alpha d)f(x, q)$, where the $f(x, q)$ is the **influence effort** and $a_i$ is the **influence factor,** which might vary by idea, and $\alpha \in [0, 1]$ represents how differentially difficult it is to *weaken* an idea. If $\alpha = 1$, weakening an idea is impossible, while if $\alpha = 0$ weakening an idea is no more difficult than strengthening it. In principle, $\alpha < 0$ could represent it being relatively easy to weaken idea, although we do not think of that case as common. The functional form of $f(x, q)$ is common across ideas (but might vary across topics or influence actors). It is increasing in both its arguments, but with diminishing marginal products and decreasing returns to scale. Finally, we'll assume that the marginal return to quality decreases more slowly than the return to quantity, such that the expansion path of influence involves increasing use of quality-intensive production as the level of influence increases.

Each idea is endowed with some base strength $0 \leq b_i$, which represents the strength of that idea in the absence of any intervention by the influence actor. Investments in influence increase the **strength** of each idea in the discourse, for investments $(x_i, q_i)$, the strength of idea i is given by $s_i = max\{b_i + a_i(d + \alpha d^2)f(x_i, q_i), 0\}$ , and its equilibrium **prominence** $(z_i)$ is given by

$$z_i = \frac{s_i}{s_1 + s_2 + \cdots + s_N}, \quad (1)$$

where we chose the typical Tullock/market-share form for this contest function for computational simplicity, although it limits the substitution patterns among the ideas and would probably need to be generalized for empirical estimation (Berry and Haile, 2021).

Finally, the influence actor is endowed with preferences that depend on the equilibrium prominence of each idea and on the total costs of the investments they make. Specifically, assume their preferences are given by

$$u(z, x, q) = \sum_{i=1}^{N} [u_i z_i - p_x x_i - p_q q_i], \quad (2)$$

where $u_i \geq 0$ is the marginal return to the prominence of idea i.

With this setup, the influence actor's optimal behavior is fairly easy to characterize. We begin by characterizing the optimal input choices, conditional on some generic strength profile and only then work back to characterizing the optimal influence profile.

*Cost-Minimizing Input Profile*

For a targeted strength profile $(s_1, s_2,..., s_N)$, the CIO's cost-minimization problem is to choose the vector of inputs (x,q) that minimizes $p_x \sum x_i + p_q \sum q_i$ while still satisfying $f(x_i, q_i) = \frac{s_i - b_i}{a_i(d + \alpha d^2)}$, for each i.

But given our assumptions about the symmetry and separability of production, this problem can be solved idea by idea, and the solution is symmetric across all ideas. For any strength profile that includes inducing strength of $s_i$ for idea i, the cost-minimizing input mix contributing to that strength, $(x^*_i, q^*_i)$, can be characterized by two equations:

$$\frac{f_x(x^*_i, q^*_i)}{f_q(x^*_i, q^*_i)} = \frac{p_x}{p_q}, \ and \qquad (3)$$

$$b_i + a_i(d + \alpha d^2)f(x^*_i, q^*_i) = s_i. \qquad (4)$$

The first equation says that the relative marginal productivities of the quality and quantity have to be equal to their relative prices. Otherwise, the input mix should be shifted to the more cost-efficient means. The second equation says that sufficient inputs need to be used to attain the desired strength.

These conditions have several immediate implications about how a CIO will produce its desired profile of idea strengths. First, if two ideas require the same level of influence effort, that effort will be produced in the same way. This is true even if (for instance) they have very different influence factors, base prominence rates, or direction of influence. Second, for given prices, the cost of producing a given level of strength is given by an increasing, convex function $c(\cdot)$, which takes as an argument the required level of influence effort, $\frac{s_i - b_i}{\alpha_i d_i(1 + \alpha_i d_i)}$.

*Optimal Narrative Strengths*

With these cost-functions in hand we can now back up to the optimal decision about idea strength. We can now rewrite the CIO's problem as

$$\max_{s: s_i \geq 0} (u_1 s_1 + u_2 s_2 + \cdots + u_N s_N)/\sum s_i \ - \ \sum c(\frac{s_i - b_i}{a_i d_i(1 + \alpha d_i)}) \qquad (5)$$

The solution to this problem breaks into two cases, where it's useful to define $S = \sum s_i$ and

$U = \sum s_i u_i$, where U/S represents the overall prominence-weighted average payoff from the entire topic.

- If $u_i - \frac{U}{S} > 0$, so the CIO would like to increase the prominence of idea i,
  - $s_i^* = b_i \; for \; all \; i \; such \; that \;\; u_i - \frac{U}{S} < \frac{S}{a_i(1+\alpha)}c'(0), \;\; and$
  - $s_i^* \; satisfying \;\; u_i - \frac{U}{S} = \frac{S}{a_i(1+\alpha)}c'(\frac{s_i^* - b_i}{a_i(1+\alpha)}), \; for \; all \; others.$
- While if $u_i - \frac{U}{S} < 0$, so the CIO would like the decrease the prominence of idea i,
  - $s_i^* = b_i \; for \; all \; i \; such \; that \;\; \frac{U}{S} - u_i < \frac{S}{a_i(1-\alpha)}c'(0),$
  - $s_i^* = 0 \; for \; all \; i \; such \; that \;\; \frac{U}{S} - u_i > \frac{S}{a_i(1-\alpha)}c'(\frac{b_i}{a_i(1-\alpha)}), \; and$
  - $s_i^* \; satisfying \;\; \frac{U}{S} - u_i = \frac{S}{a_i(1-\alpha)}c'(\frac{b_i - s_i^*}{a_i(1-\alpha)}), \; for \; all \; others.$

The left-hand side of these conditions represent the degree to which the CIO prefers idea i to be prominent, relative to the prominence-weighted average idea. The right-hand side represents how difficult it is to make idea i more prominent, which increases as the overall strength of ideas in this topic (S) increases or as the cost of exerting influence ($c'()$) increases but decreases as the strength of the particular idea is more amenable to influence ($a_i$) or when the goal is weakening versus strengthening ($\alpha$).

In one natural case where the cost of the first unit of influence is very small ($c'(0) = 0$), these conditions say that the CIO will invest in *strengthening* all ideas that yield a higher marginal utility than the weighted-average received in equilibrium from all ideas and in *weakening* all ideas that yield worse marginal utility than the average. If the cost of the first unit of influence is substantial ($c'(0) > 0$), the CIO will influence only those ideas that give a utility that is *sufficiently* different from the average, and the degree it is different from the average needs to be larger if the idea is harder to influence (larger $S/a_i$).

For those idea that the CIO decides to influence, it will exert greater influence when it cares more about the idea's prominence, relative to the average narrative (higher $|u_i - U/S|$), when the idea is easier to influence (larger $a_i$), when the cost of exerting more influence is smaller (smaller $c'(\cdot) \; and \; c''(\cdot)$, perhaps from lower input prices) and when the overall strength of the topic is lower (smaller S), either because of lower base strengths ($b_i$'s) or because of all-else-equal smaller investments in other ideas by the CIO. As long as $\alpha > 0$, so strengthening is easier than weakening, the CIO will be more likely to exert influence and will exert more effort to strengthen ideas that they prefer, rather than to directly weaken the idea that they do not.

Finally, anything that leads to greater investments in influence, will induce the CIO to use more quality-intense influence production methods.

## III. B. Putting the Model to Work

Consider two simplified polar strategic situations in which a CIO might find itself. One, which we refer to as a pure promotion goal, is a strategic situation in which the CIO has one particular idea that it is especially interested in making more prominent. One, which we refer to as pure demotion goal, is a situation in which the CIO has one particular idea that it is especially interested in making less prominent.

### III. B. 1. Pure Promotion Goal

Assume $u_1 = 1$ for the focal idea and $u_i = 0$ for the (N-1>1) alternatives . The focal idea has influence coefficient $a_1$, and all the alternatives share $a_i = a$. All ideas begin as equally prominent, with $b_i = 1$.

Under these conditions, if the CIO influences the focal idea, it will do so to increase its strength to $s_{f,pro} \geq 1$. The CIO will use the direct mechanism of strengthening the focal idea whenever

$$\frac{(N-1)s_{a,pro}}{(1+(N-1)s_{a,pro})^2}a_1(1+\alpha) > c'(0), \quad (6)$$

where $s_{a,pro}$ represents the strength of the alternative ideas. If the CIO does choose to influence the focal idea, it will choose a level of influence such that the equilibrium strength of the focal idea satisfies

$$\frac{(N-1)s_{a,pro}}{(s_{f,pro}+(N-1)s_{a,pro})^2}a_1(1+\alpha) = c'(\frac{s_{f,pro}-1}{a_1(1+\alpha)}). \quad (7)$$

If the CIO influences the alternative ideas, it will only ever *lower* their strength to $s_{a,pro} \leq 1$. As all the alternative ideas are symmetric, the CIO will either lower them all or none of them. It will will use this alternative mechanism of weakening alternatives if

$$\frac{s_{f,pro}}{(s_{f,pro}+(N-1)s_{a,pro})^2}a(1-\alpha) > c'(0), \quad (8)$$

where $s_{f,pro}$ represents the strength of the focal idea. If the CIO does choose to influence the alternative ideas, it will induce equilibrium strength $s_{a,pro}$ satisfying

$$\frac{s_{f,pro}}{(s_{f,pro}+(N-1)s_{a,pro})^2}a(1-\alpha) = c'(\frac{1-s_{a,pro}}{a(1-\alpha)}).^2 \qquad (9)$$

All else equal, more influence is exerted on strengthening the focal idea than on weakening the alternative ideas. This pattern occurs for two reasons. First, the gap between the payoff from the focal idea and the average idea is larger than the gap between the alternative ideas and the average, because the average is mostly made up from the alternative ideas. That large gap is what makes the focal idea special. Second, more influence is used on the focal idea because strengthening ideas is easier than weakening them, and the CIO is interested in strengthening the focal idea but weakening the alternatives. Despite all those factors, if influence is sufficiently inexpensive and effective both strategies will be used simultaneously. If both strategies are used, we would expect more quality-intense efforts to be used in strengthening the focal idea than in weakening the alternatives.

Obviously, this general pattern can be upset if the focal idea is particular hard to influence, relative to the alternatives ($a_1 << a$). This might be the case if the focal idea is particularly implausible or boring or complex.

Increasing the number of alternative ideas has offsetting effects for the investment in the focal idea. On the one hand, it increases the gap between payoff from the focal idea and that of the average idea– making investment more attractive. On the other hand, more ideas means a stronger topic, overall, reducing the impact of strength on idea prominence– making investment less attractive. But more alternatives always makes investing in weakening alternatives less attractive. We would, therefore, expect lower investment in each alternative and lower-quality investment in the alternative ideas as they are more numerous.

To sum up, when facing a situation in which there is one idea that the CIO would like to make more prominent, it will almost always use the direct mechanism of investing in strengthening that idea but will sometimes sometimes simultaneously engage in the mechanism of weakening alternatives– if there aren't too many of them and weakening is sufficiently effective. Only in the rare extreme case where the focal idea is very difficult to strengthen might we see the CIO exclusively using the alternative mechanism.

This case covers the first two motivational examples, the Russian promotion of certain ideas about the U.S. by strengthening those narratives and their promotion of certain ideas about Russian successes in Ukraine by the weakening of alternative ideas. According to the model, the choice between those two paths to influence is driven, primarily, by the difficulty of strengthening the two sorts of narratives. Usually, the more direct path of targeting the focal narrative is the more efficient approach, but when that idea is implausible, boring, or very unpopular, or when the alternatives are plausible, interesting, or attractive, the indirect path may dominate. Here, the ideas pushed by the IRA in America were very easy to strengthen– they had natural audiences and fit with important organic conversations around race and culture that were already growing in prominence. In contrast, the idea that Russia was doing well in Ukraine was very difficult to push. There were no punchy stories of quick victories– in contrast to initial expectations, the invasion was clearly bogging down. Also, the story of a large country defeating its smaller neighbor is just not interesting relative to the David versus Goliath story Ukraine could tell. The alternative underdog story is much more likely to go viral. But, refocusing influence efforts to undermine alternative narratives could, and likely did, prove appealing. Fact checks of videos of Ukrainian successes are interesting because Ukraine and its allies creating fake videos would be big news and fit with broader narratives regarding the ubiquity of disinformation on social media (a narrative made famous by Russia itself). Outside the mainstream media, these fact checks would be very attractive to the conspiratorially minded and debunks of these fake fact-checks would be unlikely to convince them. Furthermore, those sort of meta-debunks might just confuse people. When you are behind on the facts, muddying the waters might be the only plausible strategy.

---

[2] The influence could potentially reduce the strength of these alternative ideas down to zero, but we ignore that boundary case, here.

*III. B. 2. Pure Demotion Goal*

Assume $u_1 = 0$ for the focal idea and $u_i = 1$ for the (N-1>1) alternatives. The focal idea has influence coefficient $a_1$, and all the alternatives share $a_i = a$. All ideas begin equally prominent with $b_i = 1$.

Under these conditions, the CIO will use the direct mechanism and exert influence to weaken the focal idea if

$$\frac{(N-1)s_{a,dem}}{(1+(N-1)s_{a,dem})^2}a_1(1-\alpha) > c'(0). \qquad (11)$$

If it does exert influence to weaken this focal idea, the equilibrium strength, $s_{f,dem}$, will be chosen to satisfy

$$\frac{(N-1)s_{a,dem}}{(s_{f,dem}+(N-1)s_{a,dem})^2}a_1(1-\alpha) = c'(\frac{1-s_{f,dem}}{a_1(1-\alpha)}).^3 \qquad (12)$$

The CIO may also use the indirect mechanism and exert influence on strengthening the alternative ideas, by symmetrically choosing $s_{a,dem} \geq 1$. It will do so whenever

$$\frac{s_{f,dem}}{(s_{f,dem}+(N-1))^2}a(1+\alpha) > c'(0), \qquad (13)$$

where $s_{f,dem}$ is the strength of the focal idea, and if the CIO does exert influence in this way, $s_{a,dem}$ will be chosen to satisfy

$$\frac{s_{f,dem}}{(s_{f,dem}+(N-1)s_{a,dem})^2}a(1+\alpha) = c'(\frac{s_{a,dem}-1}{a(1+\alpha)}). \qquad (14)$$

---

[3] Again, we ignore the boundary case weakening the narrative to zero strength.

In contrast with the promotion case, we cannot decisively say which mechanism the CIO will use first, even if all ideas are equally easy to influence ($a_1 = a$). With a pure demotion goal, each mechanism has an advantage: focusing on weakening the focal idea is attractive because its payoff diverges more from the average; but focusing on strengthening alternative ideas is also attractive, because strengthening ideas is easier than weakening them. Either can dominate, depending on their relative scales. If, for instance, weakening ideas is very difficult ($\alpha \approx 1$), the indirect mechanism will dominate. But if, for instance, weakening is no more difficult than strengthening ($\alpha \approx 0$) the direct mechanism will dominate. And, of course, the direct mechanism becomes more attractive as the focal idea become relatively easy to influence ($a_1$ increases, relative to $a$).

As before, increasing the number of alternative ideas has an ambiguous impact on investments in the focal idea, but unambiguously decreases investments in influencing the alternative ideas.

This case covers the second two motivational examples, the Chinese demotion of the specific idea of labeling the Beijing Olympics as the #GenocideGames through weakening that idea by flooding it with junk content and of their demotion of the idea of an insider (virologist Li-Meng Yan) revealing that COVID originated in a lab by strengthening alternative narratives about her motivations. As in the promotion case, the decision between these paths is mostly about the difficulty of influence. As the #GenocideGames idea was quite narrow, a specific hashtag on a specific platform, it was relatively easy to affect its strength directly– flooding with bots would likely be enough to muddy the waters and reduce its prominence. But Li-Meng Yan's claim about the COVID lab origination had already been covered on Fox News and in the New York Times. Directly weakening that idea would be costly and difficult. Instead, strengthening titillating alternatives involving bribes and anti-Asian racism proved a more efficient route.

### III. B. 3. Contrasting Promotion and Demotion

In addition to comparative-static style analyses within each of the examples, we can also usefully compare the conditions in the two examples above to learn something about how CIO behavior when the goal is primarily promotion compares and contrasts with that when the goal is primarily demotion.

All-else-equal, the highest level of influence effort is applied in the case strengthening the focal idea in promotion, and the lowest level of influence goes to weakening the alternative ideas in promotion. The two demotion cases fall between them, and either one can receive more effort than the other (as discussed in section III.B.2.). This ranking also implies that we expect to see the highest quality of influence effort used in attempts to promote important ideas, with lower quality inputs applied to the other tasks, including the lowest quality in weakening alternative ideas in the promotion case.

If we stretch our examples a little to allow for the alternative ideas to be heterogeneous with respect to their influence coefficients, then the CIO will "start" with the alternatives that are easier to affect, it will invest more in influencing those alternatives, and will use higher-quality inputs in conducting that influence. Furthermore, the contrast among those alternative investments will be stronger in the demotion case, where the actor is strengthening the alternatives than it will be in the demotion case.

## IV. Generalizing the Paths to Influence

### IV. A. Beyond Our Three Examples– The Preliminaries of Quantification

The examples that motivated the analysis in this paper were chosen to represent a variety of paths of influence, to draw the reader's attention to the heterogeneity of CIO behavior. The role of the model, in this exercise, is to formalize an intuitive account of why that heterogeneity might arise and what it might imply about how CIOs operate. To play this role, the model does not have to be taken very literally– as long as it captures the key features of the strategic situation it does not matter much if the details are rough or wrong, as these simplifications can ease exposition and highlight core intuitions.

But if we are willing to take the details of the strategic situation more seriously, at the cost of complexity, a model such as we present in Section III could take up a different role, as the backbone of an empirical estimation exercise. The payoff of that approach would be substantial. It could reveal details of the productive possibilities of various influence actors which could inform countermeasures. It could be used to estimate actual influence by simulating counterfactual prominence distributions in the absence of the CIO interference. It could even be used to infer CIO priorities– how much a given CIO values prominence of various ideas in the narrative and how those values shifted over time.

Methodologies for this sort of structural estimation in other economic settings are well developed but face substantial hurdles in this context (Gandhi & Nevo, 2021). The key hurdles relate to the availability of representative data, on both the supply and demand side. First, on the supply side, we lack a full picture of CIO output choices. Datasets are mostly limited to those provided by the social media companies themselves, which comes with a medley of selection issues. For one, the companies have complicated incentives around disclosing troll behavior. Disclosures highlight vulnerabilities to users, advertisers, and shareholders alike. Any datasets created by the platforms are inherently restricted to the trolls and CIO actors that they successfully catch – potentially biasing the samples to less sophisticated CIO tactics. Third, disclosing CIOs can become implicit feedback to future CIO actors regarding how to create more resilient and survivable accounts in the future. Fourth, not all platforms devote the same resources to detection or follow the same disclosure practices. To the extent that campaigns or sub-campaigns flow across platforms (or, perhaps more importantly, differ in that choice), analyses that depend on platform disclosures might miss important outputs on non-reporting platforms.

The data shortfall is also quite severe on the demand side. Estimating the demand side of models of the sort outlined in Section III requires, at least, the market share attained by each idea. For broadcast or print media, subscription or viewership numbers are available. Put together with transcripts, and a methodology for categorizing text into ideas, one could feasibly estimate idea shares.[4] But using this approach on social media has several complexities. First, most social-media data that are available at sufficient resolution to recognize specific ideas are on output— message counts, re-shares, reactions, and the like. But the prominence of an idea is not only (or even primarily) represented by these output numbers but, rather, by the share of the messages viewed by the relevant people which represent that idea. But social-media viewership data is rarely available at sufficient resolution to make that determination, especially over time. Second, even when viewership is available, it is almost never possible to link those data to individual viewers, or even groups of viewers. Thus, identifying viewership among the **relevant** population (as determined by the CIO's preferences), is particularly difficult.

Despite the difficulties in acquiring data on both sides of the market, none of these hurdles seems completely insurmountable. On the supply side, we could begin by reasonably limiting our attention to CIOs that focus on one or two platforms, where their activities have been well documented or to CIOs with insider leaks that have revealed the full scope of their activities. But, eventually, our field needs to eventually move beyond our full dependence on platform-specific CIO identification and releases. Cross-platform methodologies for detection and measurement must be developed. Preliminary progress on the demand side of the problem might be made in similar ways. Some platforms (Youtube) do publish views data, which could be gathered over time. Others (Twitter) have recently implemented it. There is also Neilson-style tracking data on views for small panels of users. Taken together, these two sorts of data could be used to build models to predict actual view rates, as a function of more observable features (like likes and shares). But, again, cross-platform methodologies are going to be required to do this right.

### IV. B. Beyond International Politics

All of our examples of CIO behavior, so far, have been drawn from the context of international politics, but this phenomenon, and our analysis of it, is not limited to that context. On the contrary, commercial applications of coordinated influence operations are common. To wrap up this essay, we will briefly discuss how our results can inform our understanding of three of them: financial scams, fake reviews, and reputation management.

Financial scams use CIOs to make financial assets seem more or less popular, valuable, or legitimate than they actually are. This inauthentic popularity can be used for a variety of final purposes, including simple astroturfing, pump-and-dump fraud, making balance sheets look better than they really are, or establishing individual advisors' reputations for achieving super-normal returns (Xu & Livshits, 2019; Li, Shin, & Wang, 2021). As most of these schemes have the goal of inflating valuations, our promotion example from section III.B.1. is a good starting point. According to this model, we should expect focused attention on a single asset, using relatively high-quality inputs. The strength of the investment will be bigger in smaller asset markets (where the overall strength of the topic is smaller), and in situations where influence is more impactful (sayb when astro-turfing detection technologies are weaker). We should see the targeting of alternative assets for demotion only rarely (unless shorting is feasible, which turns this more into the demotion case),  when the set of alternatives is quite small and demotion is quite easy. Xu and Livshits (2019) find, for instance, that pump-and-dump influencers target crypto exchanges with relatively low volumes and which have weak influence-detection techniques.

---

[4] See, for example, Ash, Gauthier, & Widmer (2023) on the detection of economic narratives in Congressional speeches and other texts.

A second sort of commercial behavior that could be well captured by our model of CIO behavior is the production of fake product and service reviews. There is a large literature on the detection of fake reviews, and a smaller literature on their economic impacts, but relatively little is known of the determinants of investment in those reviews (Mayzlin, D., Dover, Y., & Chevalier, J., 2014;Luca & Zervas, 2016). Again, these schemes often have the goal of making a small set of products appear to be more popular or high quality than they really are, so the promotion example from section III.B.1 is probably the best fit. According to this model, we should expect focused attention on a single asset, using relatively high-quality inputs. The strength of the investment will be bigger for markets with fewer substitute products, where the initial product strength is low, as with new products, in situations where influence is more impactful (say.. when demand for the product is more elastic to influence, as may be the case with less differentiated / more competitive products). We should see the targeting of substitutes for weakening only rarely,  when the set of alternatives is quite small and weakening is quite easy– again, pointing to situations where the products are close substitutes. Luca and Zervas (2016) find patterns in Yelp restaurant ratings that fit this prediction. There are twice as many presumably fraudulent 5-star reviews as 1-star reviews, all driven by independent (non-chain) restaurants. Restaurants with more reviews are less likely to have fraudulent 5-star reviews. Restaurants that face many chain competitors are less likely to create fake reviews, but those with close substitutes (other independents of the same food time), are more likely to do so.

A final example of commercial behavior that could be captured by our model is reputation management. Here, we might interpret the topic as the set of discourses or narratives in some population about some individual or organization. The individual might like some of those, as they burnish its reputation, but might be embarrassed or threatened by others. The motivated entity could hire a reputation management service to engage in influence activities to shift the prominence of this set of ideas. This application could easily be mostly promotion or mostly demotion, depending on the context, but perhaps the most interesting case is when the entity has one particular story or idea about them that they are particularly interested in hiding– often an embarrassing or revealing incident from their past. In this case, the demotion model from section III.B.2 is probably the best fit. That model predicts that the strategy of the reputation management firm will turn crucially on how easy weakening is, relative to strengthening. If, for instance, weakening an idea is easy– as it might be in a regime with easy filing of takedown requests or little protection against costly libel litigation– the firm will focus on that approach. But if weakening is difficult, we should expect a promotion strategy, with significant investments in the strengthening of alternative ideas to crowd out the disfavored idea. Again, these investments will be larger if the pre-intervention ideas about the topic/person were pretty weak, and if the conditions of influence make efforts more impactful (as if, for example, bad search algorithms make results easy to game). Unlike the two cases above there is very limited empirical evidence about the use of reputation management services, like these, but our model could very easily structure such an investigation.

**V. Conclusion**

This paper investigated the many paths by which a coordinated actor might affect the prominence of the alternative narratives and what might drive the choice amongst those paths. We built a formal economic model of that decision and used it to explain the observed choices in four case studies. We sketched out how you might use this model to perform a more quantitative estimation of its key parameters and how those parameters might inform both our scientific understanding of these markets and help in designing policies to govern them. Finally, we extended beyond the domain of politics to show that this model can also be useful in the analysis of commercial influence.

**Bibliography**

Ash, E., Gauthier, G., & Widmer, P. (2023). Relatio: Text Semantics Capture Political and Economic Narratives. Political Analysis, 1-18. doi:10.1017/pan.2023.8

Berry, S. & Haile, P. (2021) The Foundations of Demand Estimation. *NBER Working Paper #29305*, https://doi.org/10.3386/w29305

Chen, A. (2015, June 2). The agency. *The New York Times.*
https://www.nytimes.com/2015/06/07/magazine/the-agency.html

Cohen, B., Wells, G., & McGinty, T. (2019, October 16). How one tweet turned pro-China trolls against the NBA. *The Wall Street Journal.*
https://www.wsj.com/articles/how-one-tweet-turned-pro-china-trolls-against-the-nba-11571238943

Forbes, K. (1988) Pricing of Related Products by a Multiproduct Monopolist *Review of Industrial Organization*, *3(3)*, 55-73, https://doi.org/10.1007/BF02229566

Gandhi, A. & Nevo, A. (2021) Empirical models of demand and supply in differentiated products industries in *Handbook of Industrial Organization*, 4(1) Eds. Ho, K., Hortaçsu, A. & Lizzeri, A. Elsevier, 63-139, https://doi.org/10.1016/bs.hesind.2021.11.002.

Ehrett, C., Linvill, D. L., Smith, H., Warren, P., Bellamy, L., Moawad, M., Moran, O., & Moody, M. (2022). Inauthentic newsfeeds and agenda setting in a coordinated inauthentic information operation. *Social Science Computer Review, 40(6),* 1595-1613. https://doi.org/10.1177/08944393211019951

Li, T. and Shin, D., & Wang, B., (2021) Cryptocurrency Pump-and-Dump Schemes Available at SSRN: https://ssrn.com/abstract=3267041 or http://dx.doi.org/10.2139/ssrn.3267041

Linvill, D. L., & Warren, P. L. (2020). Troll factories: Manufacturing specialized disinformation on Twitter. *Political Communication, 37,* 447-467. https://doi.org/10.1080/10584609.2020.1718257

Luca, M. & Zervas, G. (2016) Fake It Till You Make It: Reputation, Competition, and Yelp Review Fraud. Management Science 62(12):3412-3427 https://doi.org/10.1287/mnsc.2015.2304

Lyngaas, S., & Rabinowitz, H. (2023, March 13). FBI says $10 billion lost to online fraud in 2022 as crypto investment scams surged. CNN. https://www.cnn.com/2023/03/13/politics/fbi-online-fraud-report/index.html

Martin, D. A., Shapiro, J. N., & Nedashkovskaya, M. (2019). Recent Trends in Online Foreign Influence Efforts. *Journal of Information Warfare*, *18*(3), 15–48. https://www.jstor.org/stable/26894680

Maizland, L. (2022, September 22). China's repression of Uyghurs in Xinjiang. *Council on Foreign Relations.* https://www.cfr.org/backgrounder/china-xinjiang-uyghurs-muslims-repression-genocide-human-rights

Mayzlin, D., Dover, Y., & Chevalier, J. 2014. "Promotional Reviews: An Empirical Investigation of Online Review Manipulation." American Economic Review, 104 (8): 2421-55. https://doi.org/10.1257/aer.104.8.2421

Paul, C. and Matthews,M (2016) "The Russian `Firehose of Falsehood' Propaganda Model: Why It Might Work and Options to Counter It." Santa Monica, CA: RAND Corporation. https://www.rand.org/pubs/perspectives/PE198.html. https://doi.org/10.7249/PE198

Sabbagh, D. (2022, March 10). Drone footage shows Ukrainian ambush on Russian tanks. *The Guardian.*
https://www.theguardian.com/world/2022/mar/10/drone-footage-russia-tanks-ambushed-ukraine-forces-kyiv-war

Silverman, J., & Kao, J. (2022, March 11). Infamous Russian troll farm appears to be source of anti-Ukriane propaganda. *ProPublica.*
https://www.propublica.org/article/infamous-russian-troll-farm-appears-to-be-source-of-anti-ukraine-propaganda

Silverman, J., & Kao, J. (2022, March 8). In the Ukraine conflict, fake fact-checks are being used to spread disinformation. *ProPublica.*
https://www.propublica.org/article/in-the-ukraine-conflict-fake-fact-checks-are-being-used-to-spread-disinformation

Solon, O., Simmons, K., & Perrette, A. (2021, October 21). China linked disinformation campaign blames Covid on Maine lobsters. *NBC News.*
https://www.nbcnews.com/news/china-linked-disinformation-campaign-blames-covid-maine-lobsters-rcna3236

Timberg, C. (2021, February 12). Virus claim spread faster than scientists could keep up. *The Washington Post.*
https://www.washingtonpost.com/technology/2021/02/12/china-covid-misinformation-li-meng-yan/

Wells, G., & Lin, L. (2022, February 8). Pro-China Twitter accounts flood hashtag critical of Beijing Winter Olympics. *The Wall Street Journal.*
*https://www.wsj.com/articles/pro-china-twitter-accounts-flood-hashtag-critical-of-beijing-winter-olympics-11644343870*

 Xia, Y., Lukito, J., Zhang, Y., Wells, C., Kim, S. J., & Tong, C. (2019). Disinformation, performed: self-presentation of a Russian IRA account on Twitter. *Information, Communication, & Society, 22(11),* 1646–1664. https://doi.org/10.1080/1369118X.2019.1621921

Xu, J. & Livshits, B. (2019) The Anatomy of a Cryptocurrency Pump-and-Dump Scheme *Proceedings of the 28th USENIX Security Symposium* USENIX Association,1609-1625. https://doi.org/10.48550/arXiv.1811.10109

Zhang, A. (2021, July 1). #StopAsianHate: Chinese diaspora targeted by CCP disinformation campaign. *The Strategist.*
https://www.aspistrategist.org.au/stopasianhate-chinese-diaspora-targeted-by-ccp-disinformation-campaign/